\documentclass[aps,twocolumn,nofootinbib]{revtex4-2}

\usepackage{graphicx,epsfig}
\usepackage{amsmath}
\usepackage{subfigure}
\usepackage{dcolumn}
\usepackage{amsmath}
\usepackage{amsfonts}
\usepackage{amssymb}
\usepackage{braket}
\usepackage{amssymb}
\usepackage{amsmath}

\usepackage{graphicx}
\usepackage{subfigure}
\usepackage{makeidx}
\usepackage{capt-of}
\usepackage{float}
\usepackage{placeins}
\usepackage{perpage}
\usepackage{mwe}
\usepackage{kantlipsum}
\usepackage{graphicx}
\usepackage{amsfonts}
\usepackage{xcolor}
\begin{document}

\title{Holographic warm inflation}
\author{Abolhassan Mohammadi}
   \email{a.mohammadi@uok.ac.ir; abolhassanm@gmail.com}
   \affiliation{Department of Physics, Faculty of Science, University of Kurdistan, Sanandaj, Iran.}
%    \affiliation{Physics}
%\author{Mr Physics}
%   \email{ggg@gmail.com}
%   \affiliation{Department}

\begin{abstract}
The increasing interest in studying the role of holographic dark energy in the evolution of the very early universe motivates us to study it for the scenario of warm inflation. Due to this scenario, the holographic dark energy, which now drives inflation, has an interaction with the radiation. The case of interacting dark energy also has received increasing interest in studying the late time cosmology. The Infrared cutoff is taken as the Hubble length and all corrections are assumed to be exhibited by the parameter $c$, which appears in the holographic dark energy. By comparing the predictions of the model with observational data, the free constants of the model could be determined. Then, by using these values of the constants, the energy density of inflation is estimated. Next, we consider the validity of the fundamental assumptions of the warm inflation, e.g. $T/H > 1$, which is necessary to be held during inflation, for the obtained values of the constant. Gathering all outcomes, the model could be count as a suitable candidate for warm inflation.
\end{abstract}

\date{\today}
\maketitle
%\tableofcontents

%%%%%%%%%%%%%%%%%%%%%%%%%%%%%%%%%%%%%%%%%%
%%%%%%%%%%%%%%%%%%%%%%%%%%%%%%%%%%%%%%%%%%
%%%%%%%%%%%%%%%%%%%%%%%%%%%%%%%%%%%%%%%%%%
%%%%%%%%%%%%%%%%%%%%%%%%%%%%%%%%%%%%%%%%%%\\
%%%%%%%%%%%%%%%%%%%%%%%%%%%%%%%%%%%%%%%%%%
%%%%%%%%%%%%%%%%%%%%%%%%%%%%%%%%%%%%%%%%%%
%%%%%%%%%%%%%%%%%%%%%%%%%%%%%%%%%%%%%%%%%%
%%%%%%%%%%%%%%%%%%%%%%%%%%%%%%%%%%%%%%%%%%
%%%%%%%%%%%%%%%%%%%%%%%%%%%%%%%%%%%%%%%%%%
\section{Introduction}\label{Sec_intro}
Since the first proposal of inflation \cite{starobinsky1980new,Guth:1980zm}, and its preliminary modifications \cite{albrecht1982cosmology,linde1982new,linde1983chaotic}, many different models of inflation in different frames have been introduced \cite{Barenboim:2007ii,Franche:2010yj,Unnikrishnan:2012zu,Rezazadeh:2014fwa,Saaidi:2015kaa,
Fairbairn:2002yp,Mukohyama:2002cn,Feinstein:2002aj,Padmanabhan:2002cp,Aghamohammadi:2014aca,
Spalinski:2007dv,Bessada:2009pe,Weller:2011ey,Nazavari:2016yaa,
maeda2013stability,abolhasani2014primordial,alexander2015dynamics,tirandari2017anisotropic,
maartens2000chaotic,golanbari2014brane,Mohammadi:2020ake,Mohammadi:2020ctd,
berera1995warm,berera2000warm,hall2004scalar,Sayar:2017pam,Akhtari:2017mxc,Sheikhahmadi:2019gzs,Rasheed:2020syk,
Mohammadi:2018oku,Mohammadi:2019dpu,Mohammadi:2018zkf,Mohammadi:2019qeu,Mohammadi:2020ftb}. So far, the scenario of inflation has received wide acceptance from cosmologists and it has been supported strongly by the observational data \cite{Planck:2013jfk,Ade:2015lrj,Akrami:2018odb}. \\
Based on the scenario it is assumed that the scalar field is the dominant component that drives inflation. It is called inflaton. The energy density of the scalar field includes a kinetic term and a potential one. Since the scalar field varies slowly, the potential term dominates over the kinetic term. Then, the equation of the state is about $\omega \simeq -1$, and a quasi-de Sitter expansion is occurred \cite{Linde:2000kn,Linde:2005ht,Linde:2005vy,Linde:2004kg,Riotto:2002yw,Baumann:2009ds,Weinberg:2008zzc,
Lyth:2009zz,Liddle:2000cg}. Due to this extreme expansion, all other fluids diluted rapidly so that at the end of inflation the universe is cold and almost empty of particles. Therefore, to recover the hot standard big bang, a reheating mechanism is required \cite{Abbott:1982hn,Albrecht:1982mp,Dolgov:1982th,Dolgov:1989us,Traschen:1990sw,Shtanov:1994ce,Kofman:1994rk,Kofman:1997yn,Bassett:2005xm,Allahverdi:2010xz,Amin:2014eta}. \\
In 1995, a different scenario for inflation was proposed \cite{berera1995warm}, which is known as warm inflation. In the warm inflation, it is still assumed that the inflaton is still the dominant component ant it also varies slowly, however, there are some differences. One of the main differences of warm inflation is that it assumes that there is radiation along with inflaton, so that these two have interaction during the whole time of inflation. Because of the interaction, energy transfers from infaton to the radiation, and the universe remains warm at the end of reheating. Then, it comes to the second difference which is that there is no need for the reheating mechanism in the scenario of warm inflation. The next difference is about the fluctuations. In contrast to the cold inflation\footnote{the standard inflationary scenario assumes that the scalar field is the dominant component that drives inflation. All other fluids diluted rapidly and the universe is cold at the end of inflation. This is why it is also known as "cold inflation".}, where the fluctuations are quantum type \cite{Riotto:2002yw,Baumann:2009ds,Weinberg:2008zzc,Lyth:2009zz}, in the warm inflation, we have both quantum and thermal fluctuations and the thermal fluctuations dominate as long as the condition $T > H$ is satisfied \cite{berera2000warm,hall2004scalar,Moss_2007,Graham_2009,Ramos_2013,Bastero_Gil_2011}.  \\

Inspiring from \cite{Nojiri:2019kkp}, we are going to consider the role of holographic dark energy (HDE) in the very early universe. In another word, it is assumed that inflation is derived by HDE, known as a candidate of dark energy that provides interesting results for the late time evolution of the universe \cite{Nojiri:2019skr,Nojiri:2019itp,Nojiri:2005pu,Nojiri:2020wmh} (see \cite{Li:2004rb} for a review on HDE). The HDE is given by $\rho = 3c^2 M_p^2 / L$, where $c$ is a dimensionless parameter and the length scale $L$ is known as the Infrared cutoff $L_{IR}$. One of the motivations for studying the HDE for inflation is the possibility of having large HDE due to the small length scale $L$ \cite{Nojiri:2019kkp,Oliveros:2019rnq,Chakraborty_2020,Mohammadi:2021wde}. In the present work, we will study the role of HDE in warm inflation, namely, it is assumed that there is also radiation and during inflationary times HDE and radiation interact with each other. The topic has been considered for the scenario of cold inflation, however, we could not find any literature in the frame of warm inflation.  \\

There are different choices for the length scale $L$, such as different horizons and Ricci scalar. It has been shown that the Hubble horizon for the HDE could not provide desirable results and it could not provide a suitable description for the present accelerating universe \cite{Hsu:2004ri}. Ref.\cite{Malekjani_prd_2018} reconsidered the HDE with Hubble length including a varying parameter $c$. The results were promising in which the model could properly describe the late time acceleration. It motivates us to consider the same model of dark energy density in the scenario of warm inflation. Then, the length scale is taken as the Hubble horizon, and the parameter $c$ will be assumed to vary instead of being constant. \\

The paper has been organized as follows: In Sec.\ref{warminflation}, the scenario of warm inflation briefly is reviewed. In Sec.\ref{HDE_warminflation}, the HDE is taken as the source of inflation, and the main dynamical equations are derived. Then, we derive the perturbation parameters and by comparing the results with data, the free constants are determined. Next, we consider the energy scale of inflation and also investigate the validity of the main assumptions of the model. Finally, the results with be summarized in Sec.\ref{conclusion}.

%%%%%%%%%%%%%%%%%%%%%%%%%%%%%%%%%%%%%%%%%%
%%%%%%%%%%%%%%%%%%%%%%%%%%%%%%%%%%%%%%%%%%
%%%%%%%%%%%%%%%%%%%%%%%%%%%%%%%%%%%%%%%%%%
%%%%%%%%%%%%%%%%%%%%%%%%%%%%%%%%%%%%%%%%%%
%%%%%%%%%%%%%%%%%%%%%%%%%%%%%%%%%%%%%%%%%%
%%%%%%%%%%%%%%%%%%%%%%%%%%%%%%%%%%%%%%%%%%
%%%%%%%%%%%%%%%%%%%%%%%%%%%%%%%%%%%%%%%%%%
%%%%%%%%%%%%%%%%%%%%%%%%%%%%%%%%%%%%%%%%%%
%%%%%%%%%%%%%%%%%%%%%%%%%%%%%%%%%%%%%%%%%%
\section{Brief review on warm inflation}\label{warminflation}
The main dynamical equation are two Friedmann equations
\begin{eqnarray}
H^2 & = & {1 \over 3 M_p^2} \; \Big( \rho_{inf} + \rho_r \Big), \label{friedmann01} \\
\dot{H} & = & {-1 \over 2 M_p^2} \; \Big( (\rho_{inf} + p_{inf}) + (\rho_r + p_r) \Big). \label{friedmann02}
\end{eqnarray}
and the conservation equations for each fluid as \cite{Bastero-Gil:2011rva}
\begin{eqnarray}
\dot{\rho}_{inf} + 3 H (\rho_{inf} + p_{inf}) & = & - \Gamma \; (\rho_{inf} + p_{inf}), \label{cons01} \\
\dot{\rho}_r + 3 H (\rho_r + p_r) & = & \Gamma \; (\rho_{inf} + p_{inf}), \label{cons02}
\end{eqnarray}
where the subscript "inf" stands for the fluid that drives inflation (e.g. it is $\rho_{inf} = \rho_\phi$ when scalar field is the source of inflation), and the subscript "r" stands for radiation. Also, the quantity $\Gamma$ is known as the dissipation coefficient, which could be constant, depends on temperature $T_r$ or scalar field, or depends on both temperature and scalar field.  \\
Same as the cold inflation, we have slow-roll approximations which usually describe by the slow-roll parameters. The first slow-roll parameter is defined as
\begin{equation}\label{epsilon_1}
\epsilon_1 = {- \dot{H} \over H^2},
\end{equation}
and the next slow-roll parameters are defined through a hierarchy relation as follows
\begin{equation}\label{epsilon_n}
\epsilon_{n+1} = {\dot{\epsilon}_n \over H \epsilon_n}.
\end{equation}
Also, there is another type of the slow-roll parameter in warm inflation which is given by
\begin{equation}\label{SRP_beta}
\beta = {\dot{\Gamma} \over H \Gamma}.
\end{equation}
This parameter describes the evolution of the dissipation coefficient during the inflationary time.  \\

The amount of expansion of the universe during inflation is measured through the parameter $N$, known as the number of e-folds, which is defined by
\begin{equation}\label{efold}
N = \int_{t_\star}^{t_e} H dt,
\end{equation}
in which the subscripts $e$ and $\star$ respectively indicate the end of inflation and the time of horizon crossing. Using this relation, one could relate a parameter at the initial time to its value at the end of inflation.  \\

As it was mentioned before, there are both quantum and thermal fluctuations in the scenario of warm inflation, and the thermal fluctuations dominate as long as $T > H$ \cite{berera2000warm,hall2004scalar,Moss_2007,Graham_2009,Ramos_2013,Bastero_Gil_2011,Bastero-Gil:2016qru,Berera:2018tfc}. The amplitude of the scalar perturbations is given by \cite{Bastero-Gil:2016qru,Berera:2018tfc}
\begin{equation}\label{ps}
\mathcal{P}_s = {H^2 \over 8\pi^2 M_p^2 \epsilon_1} \;
\left[ 1 + 2n_{BE} + {2\sqrt{3} \pi Q \over \sqrt{3 + 4\pi Q}} \; {T \over H} \right] \; G(Q),
\end{equation}
where $n_{BE}$ is the Bose-Einstein distribution given
by $n_{BE} = \left( \exp\big( H/T_{inf} \big) - 1 \right)^{-1}$ where $T_{inf}$ is the inflaton fluctuation which is not required to necessarily be equal to radiation temperature $T_r$. Also, $G(Q)$ is a function of the dissipative parameter $Q$, given as \cite{Bastero-Gil:2016qru,Berera:2018tfc}
\begin{equation}
G(Q) = 1 + 0.0185 Q^{2.315} + 0.335 Q^{1.364}.
\end{equation}
The scalar spectral index is defined through the amplitude of the scalar perturbation as
\begin{equation}\label{ns}
n_s - 1 = {d \ln(\mathcal{P}_s) \over d \ln(k)}.
\end{equation}
The observational data determines that the scalar spectral index should be $n_s = 0.9642 \pm 0.0042$ \cite{Akrami:2018odb}, which is very close to one. Note that, $n_s = 1$ corresponds to scale-invariant fluctuations (see \cite{Weinberg:2008zzc,Lyth:2009zz,Liddle:2000cg} for more detail). \\
The amplitude of the tensor perturbation is read as \cite{Bastero-Gil:2016qru,Berera:2018tfc}
\begin{equation}\label{pt}
\mathcal{P}_t = {2 H^2 \over \pi^2 M_p^2}.
\end{equation}
The next perturbation parameter, which is widely used to test the inflationary model, is the tensor-to-scalar ratio $r$, defined by
\begin{equation}\label{r}
r = {\mathcal{P}_t \over \mathcal{P}_s}.
\end{equation}
There is still no exact data for the parameter, and the latest observational data only indicates an upper limit for the parameter as $r < 0.064$ \cite{Akrami:2018odb}.

%%%%%%%%%%%%%%%%%%%%%%%%%%%%%%%%%%%%%%%%%%
%%%%%%%%%%%%%%%%%%%%%%%%%%%%%%%%%%%%%%%%%%
%%%%%%%%%%%%%%%%%%%%%%%%%%%%%%%%%%%%%%%%%%
%%%%%%%%%%%%%%%%%%%%%%%%%%%%%%%%%%%%%%%%%%
%%%%%%%%%%%%%%%%%%%%%%%%%%%%%%%%%%%%%%%%%%
%%%%%%%%%%%%%%%%%%%%%%%%%%%%%%%%%%%%%%%%%%
%%%%%%%%%%%%%%%%%%%%%%%%%%%%%%%%%%%%%%%%%%
%%%%%%%%%%%%%%%%%%%%%%%%%%%%%%%%%%%%%%%%%%
%%%%%%%%%%%%%%%%%%%%%%%%%%%%%%%%%%%%%%%%%%
\section{HDE for warm inflation}\label{HDE_warminflation}
In this section, it is assumed that HDE is source of inflation, i.e. $\rho_{inf} = \rho_{HDE}$. The HDE is given by
\begin{equation}\label{HDE_basic}
\rho_{HDE} = {3 c^2 M_p^2 \over L^2}
\end{equation}
where $L$ is the Infrared cutoff, and $c$ is a dimensionless parameter which usually is taken as a constant, however, it could vary in a general case. \\
Here, the infrared cutoff is taken as the Hubble length, i.e. $L = H^{-1}$, and the parameter $c$ is assumed to vary instead of being constant. Such a case of HDE is studied for the late time behavior of the universe in \cite{Malekjani_prd_2018} which led to interesting results. On the other hand, it is assumed that since inflation occurs in high energy regime, there is an Ultraviolet correction to the Infrared cut off. The presence of such a corrections also assumed to be included in the parameter $c$. Then, for the Friedmann equation we have
\begin{equation}\label{friedmannHDE}
H^2 = {1 \over 3M_p^2} \left( {3 c^2 M_p^2 H^2} + \rho_r \right).
\end{equation}
From the radiation conservation equation, Eq.\eqref{cons02}, and by imposing the quasi-stable production of the radiation, i.e. $\dot{\rho} \ll H \rho_r, \Gamma \; (\rho_{inf}+ p_{inf})$, one arrives at
\begin{equation}
4 H \rho_r = \Gamma \; (\rho_{HDE}+ p_{HDE}).
\end{equation}
and by using the second Friedmann equation, Eq.\eqref{friedmann02}, the radiation energy density is read as
\begin{equation}\label{radiation_rho}
\rho_r = {-3 M_p^2 \over 2} \; {Q \over 1+Q} \; \dot{H},
\end{equation}
note that, since energy density is positive, the time derivative of the Hubble parameter should be negative. The quantity $Q$ is known as the dissipative parameter defined as $Q \equiv \Gamma / 3H$. \\
Substituting Eq.\eqref{radiation_rho}, in the Friedmann equation Eq.\eqref{friedmannHDE}, the time derivative of the Hubble parameter is obtained as
\begin{equation}\label{dHt}
\dot{H} = -2 (1 - c^2) \; {1+Q \over Q} \; H^2.
\end{equation}
Inserting the result, in Eq.\eqref{radiation_rho}, the radiation energy density is rewritten in terms of the parameter $c$ and the Hubble parameter as
\begin{equation}\label{radiation_H}
\rho_r = 3 M_p^2 \; (1 - c^2) \; H^2.
\end{equation}
On the other hand, the radiation energy density is expressed in terms of its temperature
\begin{equation}\label{radiation_T}
\rho_r = \sigma \; T_r^4,
\end{equation}
where $\sigma$ is the Stephen-Boltzman constant given by $\sigma = \pi^2 g_{\star} / 30$, where $g_\star$ is the number of degree of freedom of radiation field. $T_r$ is the temperature of the radiation. Comparing Eqs.\eqref{radiation_H} and \eqref{radiation_T}, the temperature is obtained as follow
\begin{equation}\label{temperature}
T_r^4 = {3 M_p^2 \over \sigma} \; (1 - c^2) \; H^2
\end{equation}
From the definition of $\epsilon_1$ and using Eq.\eqref{dHt}, the parameter , in general, is read as
\begin{equation}\label{epsilon_1_HDE}
\epsilon_1 = 2 (1 - c^2) \; {1+Q \over Q}.
\end{equation}
From Eq.\eqref{epsilon_n}, the second slow-roll parameter is obtained as
\begin{equation}\label{epsilon_2_HDE}
\epsilon_2 = \eta - {1 \over 1+Q} \; \big( \beta - \epsilon_1 \big),
\end{equation}
where the new parameter $\eta$ is defined as
\begin{equation}\label{eta_HDE}
\eta = {-2 c \dot{c} \over H \; (1 - c^2)}.
\end{equation}
To go for more detail, the derivation coefficient $\Gamma$ should be introduced. The coefficient could be taken as a constant, but in a more general view it is taken as a function of the temperature $T_r$ \cite{Moss:2006gt,CID2015127,Panotopoulos:2015qwa,Mishra:2011vh,Moss:2008yb,Zhang:2009ge,Zhang:2013yr}. Then, it is taken as a power-law function of the temperature, i.e.
\begin{equation}\label{Gamma_T}
\Gamma = C_T \; T^m .
\end{equation}
Where $C_T$ is a constant. Then, using Eq.\eqref{temperature}, it is expressed in terms of the parameter $c$ and the Hubble parameter
\begin{equation}\label{Gamma_H}
\Gamma = C_T \; \left( {3 M_p^2 \over \sigma} \; (1 - c^2) \; H^2 \right)^{m/4}.
\end{equation}
The result could be utilized in Eq.\eqref{SRP_beta}, so that, the slow-roll parameter $\beta$ is simplify as
\begin{equation}\label{beta_HDE}
\beta = -2 (1 - c^2) \; {1+Q \over Q} \; {H \Gamma_{,H} \over \Gamma}
\end{equation}

%%%%%%%%%%%%%%%%%%%%%%%%%%%%%%%%%%%%%%%%%%
%%%%%%%%%%%%%%%%%%%%%%%%%%%%%%%%%%%%%%%%%%
%%%%%%%%%%%%%%%%%%%%%%%%%%%%%%%%%%%%%%%%%%
%%%%%%%%%%%%%%%%%%%%%%%%%%%%%%%%%%%%%%%%%%
\subsection{Holographic warm inflation in HDR}\label{HDE_WI_HDR}
For the rest of work, it is assumed that inflation occurs in the high dissipative regime (HDR), i.e. $Q \gg 1$. Imposing this condition on the equations, and by substituting the radiation energy density \eqref{radiation_rho} in Eq.\eqref{friedmannHDE}, the first slow-roll parameter is obtained only in terms of the parameter $c$ as
\begin{equation}\label{first_SLP}
\epsilon_1 = {- \dot{H} \over H^2} = 2 (1 - c^2).
\end{equation}
Then, utilizing the hierarchy definition of the slow-roll parameters, the second one is given by
\begin{equation}\label{second_SRP}
\epsilon_2 = {\dot{\epsilon_1} \over H \epsilon_1} = {-2 c \dot{c} \over H (1 - c^2)}
\end{equation}

The parameter $c$ is assumed to be given by $c = c_0 H^\gamma$, where $c_0$ and $\gamma$ are constants which will be determined later in a comparison with observational data. By this definition, we have
\begin{equation}
\dot{H} = -2 (1 - c_0^2 H^{2\gamma}) \; H^2.
\end{equation}
Inserting the above relation in the equation of number of e-fold \eqref{efold}, one arrives at
\begin{equation}\label{c_N}
c^2(N) = {\xi_0 \; e^{-4 \gamma N} \over 1 + \xi_0 \; e^{-4\gamma N}},
\end{equation}
and from the definition of $c$, the Hubble parameter is extracted as
\begin{equation}\label{H_N}
H^{2\gamma}(N) = {\xi_0 \; e^{-4 \gamma N} \over c_0^2 \Big(1 + \xi_0 \; e^{-4\gamma N} \Big)}.
\end{equation}
Since, both the slow-roll parameters were expressed in terms of the parameter $c$, they could also be rewritten in terms of the number of e-folds, as follows
\begin{eqnarray}\label{SRP_N}
\epsilon_1(N) & = & {2 \over 1 + \xi_0 \; e^{-4\gamma N}}, \\
\epsilon_2(N) & = & {4 \gamma \; \xi_0 \; e^{-4 \gamma N} \over 1 + \xi_0 \; e^{-4\gamma N}} .
\end{eqnarray}
The next slow-roll parameter is given by
\begin{equation}\label{beta_N}
\beta(N) = -m \; \left[ 1 - (1+\gamma) \; {\xi_0 \; e^{-4 \gamma N} \over 1 + \xi_0 \; e^{-4\gamma N}} \right].
\end{equation}
By computing Eq.\eqref{ns} for the HDR, the scalar spectral index is obtained as\footnote{Note that, since we are in HDR, the dissipative parameter $Q$ is bigger than one, i.e. $Q \gg 1$. Therefore, the function $G(Q)$, in the amplitude of the scalar perturbations \eqref{ps}, is approximated as $G(Q) \simeq 0.0185 Q^{2.315}$. } 
\begin{equation}\label{ns_HDE_HDR}
ns - 1 = 1.815 \epsilon_1 - \epsilon_2 + 3.815 \beta,
\end{equation}
and the tensor-to-scalar ratio is acquired from Eqs.\eqref{ps} and \eqref{pt} as follows 
\begin{equation}\label{r_HDE_HDR}
r = 16 \; \epsilon_1 \; \left( \sqrt{3 \pi} \; {T \over H} \; 0.0185 Q^{2.815} \right)^{-1}.
\end{equation}
The constant $\xi_0$ could be estimated by computing the first slow-roll parameter $\epsilon_1(N)$ at the end of inflation. The point is that, at the end of inflation, $\epsilon_1(N)$ reaches one, i.e. $\epsilon_1 = 1$, which indicates the end of accelerated expansion phase. Then, to have $\epsilon_1(N=0) = 1$, the constant $\xi_0$ should be $\xi_0 = 1$. \\

%%%%%%%%%%%%%%%%%%%%%%%%%%%%%%%%%%%%%%%%%%
%%%%%%%%%%%%%%%%%%%%%%%%%%%%%%%%%%%%%%%%%%
%%%%%%%%%%%%%%%%%%%%%%%%%%%%%%%%%%%%%%%%%%
%%%%%%%%%%%%%%%%%%%%%%%%%%%%%%%%%%%%%%%%%%
\subsection{Comparing the model with data}\label{HDE_WI_HDR_data}
\begin{figure}[h]
\centering
\includegraphics[width=8cm]{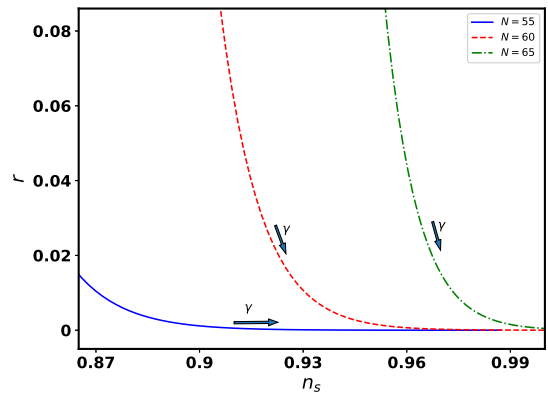}
\caption{\label{rns_gamma} The figure illustrates the tensor-to-scalar ratio $r$ versus the scalar spectral index $n_s$ for different values of number of e-fold. Here, $\gamma$ is taken as the variable and the arrow for each curve shows the direction of increasing $\gamma$. The figure shows the values of $n_s$ and $r$ at the time of the horizon crossing.}
\end{figure}

To verify the validity of the model, its results should be compared with observational data, or one could apply the observational data and constrain the free constants of the model. In this regard, first we need to express the perturbation parameters $\mathcal{P}_s$, $n_s$ and $r$ in terms of the number of e-fold. This way, the parameters could be easily estimated at the time of the horizon crossing. \\
using Eqs.\eqref{SRP_N} and \eqref{beta_N}, the scalar spectral index is expressed in terms of the number of e-fold. Next, through the Eqs.\eqref{ps}, \eqref{temperature}, \eqref{Gamma_H}, and \eqref{H_N}, it is achieved that
\begin{equation}\label{r_paranthesis}
\left( \sqrt{3 \pi} \; {T \over H} \; 0.0185 Q^{2.815} \right) \Big|_{\star} = {8\pi^2 M_p^2 \mathcal{P}_s^\star \epsilon_1(N) \over H^2(N)}.
\end{equation}
Then, substituting Eq.\eqref{r_paranthesis} in \eqref{r_HDE_HDR}, the tensor-to-scalar ratio also is read in terms of the number of e-fold. Fig.\ref{rns_gamma} illustrates the tensor-to-scalar ratio $r$ versus the scalar spectral index $n_s$, where the varying parameter is $\gamma$ at the time of the horizon crossing. The curves are plotted for different values of the number of e-fold which are mostly used in the literature. The latest data states that $n_s = 0.9642 \pm 0.0049$ and $r \leq 0.64$. By increasing the number of e-fold, the curves goes out of the observational region. However, for the smaller number of e-fold, the tensor-to-scalar ratio $r$ gets smaller values. Moreover, by taking smaller $N$, the curve stay in the observational range for bigger values of $\gamma$.
%%%%%%%%%%%%%%%%%%%%%%%%%%%%%%%%%%%5
%%%%%%%%%%%%%%%%%%%%%%%%%%%%%%%%%%%5
Some numerical results are presented in Table.\ref{numerical_rns}, which could gives more insight. It is seen that for a specific value of $\gamma$, both $n_s$ and $r$ increase by enhancement of the number of e-fold. One the other hand, for specific value of $N$, $n_s$ increases and $r$ decreases by growing of $\gamma$. More results are presented in Table.\ref{Table_numericalresults}. \\
%%%%%%%%%%%%%%%%%%%%%%%%%%%%%%%%%%%%%%
\begin{table}
\caption{\label{numerical_rns} The table shows numerical results for the scalar spectral index and the tensor-to-scalar ratio for different values of the number of e-fold $N$ and $\gamma$. The other constants are taken as $c_0 = 0.62$ and $m=3$.}
\begin{tabular}{p{1cm}p{1.2cm}p{1.8cm}p{2cm}}
\hline
$N$  & $\quad \gamma$ &  $\quad n_s$  &  $\quad r$   \\
\hline
\hline
$55$   &  $0.015$  &  $0.8719$  &  $0.0090$   \\

$55$   &  $0.016$  &  $0.9170$  &  $3.05 \times 10^{-4}$   \\

$55$   &  $0.017$  &  $0.9555$  &  $1.41 \times 10^{-5}$   \\
\hline

$60$   &  $0.015$  &  $0.9248$  &  $0.0168$   \\

$60$   &  $0.016$  &  $0.9629$  &  $5.00 \times 10^{-4}$   \\

$60$   &  $0.017$  &  $0.9943$  &  $2.09 \times 10^{-5}$   \\
\hline

$65$   &  $0.015$  &  $0.9646$  &  $0.0267$   \\

$65$   &  $0.016$  &  $0.9964$  &  $7.18 \times 10^{-4}$   \\

$65$   &  $0.017$  &  $1.022$  &  $2.78 \times 10^{-5}$    \\
\hline
\hline
\end{tabular}
\end{table}
%%%%%%%%%%%%%%%%%%%%%%%%%%%%%%%%%%%%%%

\begin{figure}[h]
\centering
\includegraphics[width=8cm]{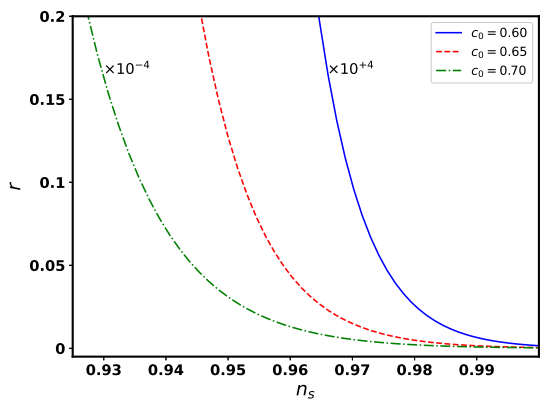}
\caption{\label{rns_c0} The plot is a parametric plot displaying the tensor-to-scalar ratio, $r$, versus the scalar spectral index, $n_s$, for different values of $c_0$. Same as Fig.\ref{rns_gamma}, the variable is $\gamma$. It is realized that by decreasing of the $c_0$ the curves goes out of the observational range, however, by enhancement of $c_0$, it comes inside and the parameter $r$ becomes very small.}
\end{figure}

Fig.\ref{rns_c0} exhibits the same plot for the different values of the constant $c_0$. The constant $c_0$ appears in the tensor-to-scalar ratio. The figure displays that by enhancement of $c_0$, $r$ dramatically decreases.\\
%%%%%%%%%%%%%%%%%%%%%%%%%%%%%%%%%%%%%%%%%%5
%%%%%%%%%%%%%%%%%%%%%%%%%%%%%%%%%%%%%%%%5

\begin{figure}[h]
\centering
\includegraphics[width=8cm]{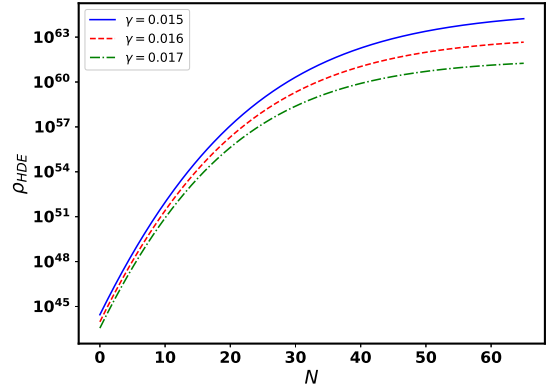}
\caption{\label{HDE_energy} The behavior of the HDE during the inflationary times is presented for different values of $\gamma$. It has a high value at the initial time and then it decreases by passing the time. It is seen that the energy density of inflation is about $10^{64} {\rm GeV^4}$. The plot also indicates that HDE has smaller values for bigger values of $\gamma$.}
\end{figure}

By comparing the model with observational data, we are provided with a general view about the values of the constants of the model to have an agreement with observation. Taking these values of the constants, we could have a general insight about the energy density $\rho_H$, which displays the energy density of inflation. Fig.\ref{HDE_energy} portrays the behavior of the HDE during the inflationary time. It is concluded that the inflation starts at the energy scale about $10^{64} {\rm GeV^4}$, and it deceases by approaching to the end of inflation. \\
%%%%%%%%%%%%%%%%%%%%%%%%%%%%%%%%%%%%%%%%%%%%%%%%
%%%%%%%%%%%%%%%%%%%%%%%%%%%%%%%%%%%%%%%%%%%%%%

%%%%%%%%%%%%%%%%%%%%%%%%%%%%%%%%%%%%%%%%%%
%%%%%%%%%%%%%%%%%%%%%%%%%%%%%%%%%%%%%%%%%%
%%%%%%%%%%%%%%%%%%%%%%%%%%%%%%%%%%%%%%%%%%
%%%%%%%%%%%%%%%%%%%%%%%%%%%%%%%%%%%%%%%%%%
\subsection{Verifying the conditions of the model}\label{model_conditions}
\begin{figure}[h]
\centering
\includegraphics[width=8cm]{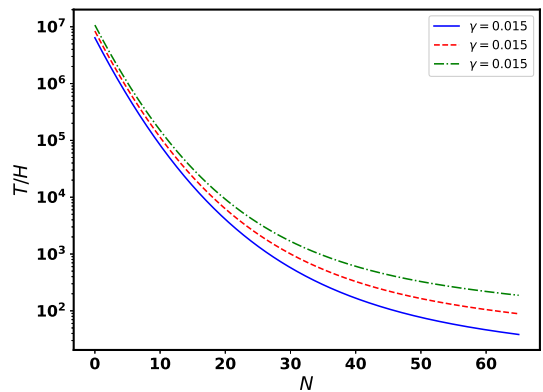}
\caption{\label{TH_N} The term $T/H$ is plotted versus the number of e-fold for different values of the constant $\gamma$. The curves shows the value of the term during the inflationary time. It is seen that the condition $T/H > 1$ is verified for the whole time of inflation. }
\end{figure}

\begin{figure}[h]
\centering
\includegraphics[width=8cm]{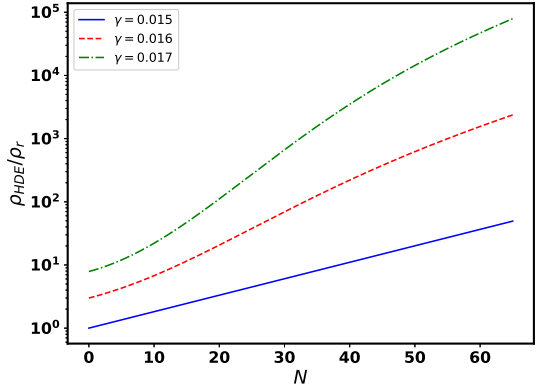}
\caption{\label{densities} The plot illustrates the ratio of energy densities of the HDE and radiation during the inflationary times. At the initial time HDE is much bigger, and by passing the time and approaching to the end of inflation, they come close together.}
\end{figure}

In the scenario of warm inflation, we have two fundamental assumptions as $T/H > 1$ and $\rho_{HDE} / \rho_r > 1$. These assumptions should be verified for the obtained values of the constants. Fig.\ref{TH_N} displays the term $T/H$ versus the number of e-fold for different values of $\gamma$. It is verified that the term is greater than one, and the condition $T/H > 1$ is verified, and also it increase by approaching to the end of inflation.
%%%%%%%%%%%%%%%%%%%%%%%%%%%%%%%%%%%%%%%%%%%%%
%%%%%%%%%%%%%%%%%%%%%%%%%%%%%%%%%%%%%%%%%%%%%%
Comparing of the energy densities is presented in Fig.\ref{densities}. The plot exhibits the ratio $\rho_{HDE}/\rho_r$ for different values of $\gamma$ during inflation. It is realized that at the initial times, the HDE is much bigger than the radiation energy density. Then, the HDE is the dominant component, and the assumption that the warm inflation is driven by HDE is verified. By the passing time, they comes close together so that at the end of inflation they are comparable to each other.
%%%%%%%%%%%%%%%%%%%%%%%%%%%%%%%%%%%5
%%%%%%%%%%%%%%%%%%%%%%%%%%%%%%%%%%%%%%%

%%%%%%%%%%%%%%%%%%%%%%%%%%%%%%%%%%%%%%%%%%
%%%%%%%%%%%%%%%%%%%%%%%%%%%%%%%%%%%%%%%%%%
%%%%%%%%%%%%%%%%%%%%%%%%%%%%%%%%%%%%%%%%%%
%%%%%%%%%%%%%%%%%%%%%%%%%%%%%%%%%%%%%%%%%%
%%%%%%%%%%%%%%%%%%%%%%%%%%%%%%%%%%%%%%%%%%
%%%%%%%%%%%%%%%%%%%%%%%%%%%%%%%%%%%%%%%%%%
%%%%%%%%%%%%%%%%%%%%%%%%%%%%%%%%%%%%%%%%%%
%%%%%%%%%%%%%%%%%%%%%%%%%%%%%%%%%%%%%%%%%%
%%%%%%%%%%%%%%%%%%%%%%%%%%%%%%%%%%%%%%%%%%
\section{Conclusion}\label{conclusion}
The scenario of warm inflation was considered by this assumption that the HDE is the source of inflation. Based on the assumptions of the scenario of warm inflation, we now have two-component as HDE and radiation. They interact with each other and the energy transfer from HDE to the radiation. Moreover, the scenario predicts two types of fluctuations as quantum and thermal fluctuation, respectively are proportional to $H$ and $T$. The thermal fluctuations dominate over the quantum fluctuation as long as the condition $T/H > 1$ is preserved. \\
The infrared of the HDE was assumed to be given by the Hubble length and we also included this assumption that the parameter $c$, which appears in the HDE, is a varying parameter instead of being constant. Then, the dynamical equations of the model and also the main perturbations parameter were derived. Next, the perturbation parameters were estimated at the time of the horizon crossing and by comparing them with observational data, we could determine the free constants of the model. The $r-n_s$ diagram of the model was depicted and it was realized that the model could come to a good agreement with data. Next, the behavior of the HDE was investigated. The HDE was assumed for the source of inflation, e.g it is the dominant component. Therefore, its energy density at the initial times gives the energy scale of inflation. The behavior of HDE was plotted which displayed that the energy scale of inflation is around $10^{64} {\rm GeV^4}$. \\
At the final step, we reconsider the verification of the fundamental conditions of the model, i.e. $T/H > 1$ and $\rho_{HDE}/\rho_r \gg 1$, for the determined values of the constants. The first condition guarantees that the thermal fluctuations dominate over the quantum fluctuations. The behavior of the term $T/H$ was plotted in Fig.\ref{TH_N}. It shows that the term is greater than one for the whole time of inflation. Also, it gets larger over time. The ratio $\rho_{HDE}/\rho_r \gg 1$ was illustrated in Fig.\ref{densities}, which determines that at the initial time, the HDE is much bigger than one. This result verifies the assumption that inflation is driven by the HDE. By approaching the end of inflation, the ratio gets smaller, stating that the two densities come closer. A numerical result of the model is presented in Table.\ref{Table_numericalresults}.

%%%%%%%%%%%%%%%%%%%%%%%%%%%%%%%%%%%%%%%%%%%5
\begin{widetext}

\begin{table}[h]
\centering
\caption{\label{Table_numericalresults} The table shows the numerical results of the model for different values of the constants.}
\begin{tabular}{p{1cm}p{1.2cm}p{1.2cm}p{1cm}p{1.5cm}p{2cm}p{2cm}p{1.5cm}p{1.5cm}}
\hline
$N$  & $\quad \gamma$ &  $\ c_0$  &  $m$  &  $\quad n_s$  &  $\quad r$  &  energy scale  &  $T/H$  &  $\rho_H / \rho_r$ \\
\hline
$55$   &  $0.015$  &  $0.62$  &  $3$  &  $0.8719$  &  $0.0090$  &  $8.68 \times 10^{16}$  &
$58.09$  &  $27.11$  \\

$55$   &  $0.016$  &  $0.62$  &  $3$  &  $0.9170$  &  $3.05 \times 10^{-4}$  &  $3.72 \times 10^{15}$  &
$128.62$  &  $33.78$  \\

$55$   &  $0.017$  &  $0.62$  &  $3$  &  $0.9555$  &  $1.41 \times 10^{-5}$  &  $1.73 \times 10^{15}$  &
$262.83$  &  $24.09$  \\
\hline

$60$   &  $0.015$  &  $0.62$  &  $3$  &  $0.9248$  &  $0.0168$  &  $1.01 \times 10^{16}$  &
$46.29$  &  $36.59$  \\

$60$   &  $0.016$  &  $0.62$  &  $3$  &  $0.9629$  &  $5.00 \times 10^{-4}$  &  $4.22 \times 10^{15}$  &
$105.14$  &  $46.52$  \\

$60$   &  $0.017$  &  $0.62$  &  $3$  &  $0.9943$  &  $2.09 \times 10^{-5}$  &  $1.91 \times 10^{16}$  &
$219.09$  &  $59.14$  \\
\hline

$65$   &  $0.015$  &  $0.62$  &  $3$  &  $0.9646$  &  $0.0267$  &  $1.14 \times 10^{16}$  &
$38.34$  &  $49.40$  \\

$65$   &  $0.016$  &  $0.62$  &  $3$  &  $0.9964$  &  $7.18 \times 10^{-4}$  &  $4.62 \times 10^{15}$  &
$88.81$  &  $64.07$  \\

$65$   &  $0.017$  &  $0.62$  &  $3$  &  $1.022$  &  $2.78 \times 10^{-5}$  &  $2.05 \times 10^{15}$  &
$38.34$  &  $49.40$  \\
\hline
\hline

$55$   &  $0.007$  &  $0.73$  &  $1$  &  $0.9601$  &  $0.0211$  &  $1.03 \times 10^{16}$  &
$70.18$  &  $4.66$  \\

$55$   &  $0.007$  &  $0.73$  &  $1$  &  $0.9715$  &  $7.53 \times 10^{-4}$  &  $4.52\times 10^{15}$  &
$154.29$  &  $5.8$  \\

$55$   &  $0.007$  &  $0.73$  &  $1$  &  $0.9761$  &  $2.86 \times 10^{-5}$  &  $2.01 \times 10^{15}$  &
$333.07$  &  $7.2$  \\
\hline

$60$   &  $0.007$  &  $0.73$  &  $1$  &  $0.9698$  &  $0.593$  &  $2.38 \times 10^{16}$  &
$29.62$  &  $5.36$  \\

$60$   &  $0.007$  &  $0.73$  &  $1$  &  $0.9750$  &  $0.0117$  &  $9.02\times 10^{15}$  &
$75.08$  &  $6.82$  \\

$60$   &  $0.007$  &  $0.73$  &  $1$  &  $0.9793$  &  $2.70 \times 10^{-4}$  &  $3.54 \times 10^{15}$  &
$182.66$  &  $8.67$  \\
\hline

$65$   &  $0.007$  &  $0.73$  &  $1$  &  $0.9730$  &  $11.47$  &  $65.03 \times 10^{16}$  &
$13.71$  &  $6.17$  \\

$65$   &  $0.007$  &  $0.73$  &  $1$  &  $0.9781$  &  $0.12$  &  $1.64 \times 10^{16}$  &
$39.92$  &  $8.00$  \\

$65$   &  $0.007$  &  $0.73$  &  $1$  &  $0.9822$  &  $0.0018$  &  $5.73 \times 10^{15}$  &
$108.82$  &  $10.38$  \\
\hline
\hline

$55$   &  $0.0170$  &  $0.58$  &  $-1$  &  $1.0297$  &  $0.03614$  &  $1.23 \times 10^{16}$  &
$36.96$  &  $42.09$  \\

$55$   &  $0.0180$  &  $0.58$  &  $-1$  &  $0.9991$  &  $0.0014$  &  $5.49 \times 10^{15}$  &
$78.51$  &  $52.45$  \\

$55$   &  $0.0195$  &  $0.58$  &  $-1$  &  $0.9617$  &  $1.93 \times 10^{-5}$  &  $1.87 \times 10^{15}$  &
$212.35$  &  $72.96$  \\
\hline

$60$   &  $0.0170$  &  $0.58$  &  $-1$  &  $0.9933$  &  $0.0536$  &  $1.36 \times 10^{16}$  &
$30.81$  &  $59.14$  \\

$60$   &  $0.0180$  &  $0.58$  &  $-1$  &  $0.9681$  &  $0.0019$  &  $5.95 \times 10^{15}$  &
$66.41$  &  $75.18$  \\

$60$   &  $0.0195$  &  $0.58$  &  $-1$  &  $0.9377$  &  $2.42 \times 10^{-5}$  &  $1.98 \times 10^{15}$  &
$182.32$  &  $107.77$  \\
\hline

$65$   &  $0.0170$  &  $0.58$  &  $-1$  &  $0.9670$  &  $0.0711$  &  $1.46 \times 10^{16}$  &
$26.40$  &  $83.09$  \\

$65$   &  $0.0180$  &  $0.58$  &  $-1$  &  $0.9463$  &  $0.0024$  &  $6.29 \times 10^{15}$  &
$57.49$  &  $107.77$  \\

$65$   &  $0.0195$  &  $0.58$  &  $-1$  &  $0.9213$  &  $2.81 \times 10^{-5}$  &  $2.06 \times 10^{15}$  &
$159.31$  &  $159.17$  \\
\hline
\hline

\end{tabular}
\end{table}

\end{widetext}
%%%%%%%%%%%%%%%%%%%%%%%%%%%%%%%%%%%%%%%%%%%

\section*{Acknowledgments}
The work of A. Mohammadi has been supported financially by
“Vice Chancellorship of Research and Technology, University of Kurdistan” under research Project No. 99/
11/19063.

\bibliography{HDE_WI}

%\begin{thebibliography}
%\bibitem{Rasheed:2020syk} A.~Mohammadi, T.~Golanbari, H.~Sheikhahmadi,
%K.~Sayar, L.~Akhtari, M.~A.~Rasheed and K.~Saaidi, ``Warm Tachyon Inflation
%and Swampland Criteria,'' Chinese Physics C \textbf{\ 44}, No. 9 (2020)
%095101, doi:10.1088/1674-1137/44/9/095101 [arXiv:2001.10042 [gr-qc]].
%2 citations counted in INSPIRE as of 11 Jul 2020
%\end{thebibliography}

\end{document}